\documentclass[twocolumn,           
               showpacs,            
               preprintnumbers,     
               aps,                 
               prd,                 
               a4paper,             
               groupedaddress,      
               nofootinbib,         
               tightenlines,        
               floats,floatfix      
               ]{revtex4}

\usepackage{graphicx}
\usepackage{subfigure}
\usepackage{latexsym}
\usepackage{amsmath,amssymb}        
\usepackage[draft=false]{hyperref}

\input colordvi

\begin{document}



\title{Model selection forecasts for the spectral index from the
Planck satellite}  
\author{C\'edric Pahud}
\affiliation{Astronomy Centre, University of Sussex, Brighton BN1 9QH,
United Kingdom}
\author{Andrew R.~Liddle}
\affiliation{Astronomy Centre, University of Sussex, Brighton BN1 9QH,
United Kingdom}
\author{Pia Mukherjee}
\affiliation{Astronomy Centre, University of Sussex, Brighton BN1 9QH,
United Kingdom}
\author{David Parkinson}
\affiliation{Astronomy Centre, University of Sussex, Brighton BN1 9QH,
United Kingdom}
\date{\today}
\pacs{98.80.-k \hfill astro-ph/0605004}
\preprint{astro-ph/0605004}


\begin{abstract}
The recent WMAP3 results have placed measurements of the spectral
index $n_{{\rm S}}$ in an interesting position. While parameter
estimation techniques indicate that the Harrison--Zel'dovich spectrum
$n_{{\rm S}}=1$ is strongly excluded (in the absence of tensor
perturbations), Bayesian model selection techniques reveal that the
case against $n_{{\rm S}}=1$ is not yet conclusive. In this paper, we
forecast the ability of the Planck satellite mission to use Bayesian
model selection to convincingly exclude (or favour) the
Harrison--Zel'dovich model.
\end{abstract}

\maketitle


\section{Introduction}

One of the key goals of cosmology is to probe the nature of the
primordial perturbations, for instance to seek support for the
inflationary cosmology. The simplest models of inflation predict
adiabatic gaussian density perturbations of approximately power-law
form, characterized by the spectral index $n_{{\rm S}}$, and in
addition a spectrum of gravitational wave perturbations (see
Ref.~\cite{LL} for an overview).

The ability of experiments, actual or proposed, to explore such
questions is typically framed in terms of parameter estimation, for
instance by forecasting the expected uncertainty on $n_{{\rm S}}$
given a particular assumed fiducial model. However, it has been
stressed in a number of papers recently \cite{L04,T,MPL,MPCLK} that
many of the key questions are not ones of parameter estimation, but of
{\em model selection} \cite{Jeff,MacKay,Gregory}. Model selection
problems are characterized by an uncertainty in the choice of
parameters to vary in a fit to data, rather than of the values of a
parameter set chosen by hand. The discovery of any new physical effect
in data is indicated by the need to include new parameters, possible
examples being non-zero spatial curvature, time variation of the dark
energy density, or the existence of tensor perturbations. Early
cosmological applications of this technique were given in
Ref.~\cite{ev}.

Since many of the most important questions are ones of model selection
rather than parameter estimation, it follows that the capabilities of
experiments should also be quantified by model selection criteria
rather than parameter uncertainty forecasts alone. This is the
viewpoint adopted in recent papers by Trotta \cite{T}, whose {\em
Expected Posterior Odds} (ExPO) forecasting technique estimates the
probability of new data requiring new parameters, and by Mukherjee et
al.~\cite{MPCLK} who use Bayes factor plots to compare the ability of
different experiments to decisively select between models.

Mukherjee et al.~\cite{MPCLK} illustrated model selection forecasting
using dark energy surveys, looking at a two-parameter dark energy
model versus a cosmological constant model. The same general approach
is applicable in many other contexts. In this paper, we carry out
model selection forecasting for the Planck satellite cosmic microwave
background project, focussing on its ability to measure the spectral
index $n_{{\rm S}}$. This is particularly timely as the recent release
of the three-year WMAP data \cite{wmap3} has placed this parameter in
the zone around three-sigma where the application of model selection
techniques is at its most crucial \cite{T}. In a companion paper to
this one, Parkinson et al.~\cite{PML} have shown that the case for
$n_{{\rm S}} \neq 1$ is far from decisive at present.

We assume throughout that there are no tensor perturbations. While it
would be interesting to explore models including tensors, thus
properly probing the inflationary space, at present doing $n_{{\rm
S}}$ alone stretches our supercomputer resources to their limit. In
this regard model selection forecasting is much more challenging than
analysis of real data, as instead of having a single dataset to
analyze, one has to create and analyze simulated datasets for a range
of possible models and model parameters.

\section{Model selection forecasts for $n_{{\rm S}}$}

\subsection{Model selection forecasting}

The philosophical underpinning of model selection forecasting was
described in Mukherjee et al.~\cite{MPCLK} and we summarize it only
very briefly here. Given a particular dataset, simulated or real,
model selection is carried out by evaluation of a model selection
statistic for each model, where the term {\em model} refers to a
choice of parameters to be varied plus a set of prior ranges for those
parameters. The usual statistic of choice is the {\em Bayesian
evidence} $E$, also known as the marginalized likelihood. The ratio of
evidences between two models is known as the Bayes factor, $B_{10} =
E(M_1)/E(M_0)$, where $M_1$ and $M_0$ indicate the two models under
consideration. By plotting the Bayes factor using datasets generated
as a function of a parameter of interest, one uncovers the regions of
parameter space in which a given experiment would be able to
decisively select between the two models, and also those regions where
the comparison would be inconclusive.

In short, the advantages of model selection over parameter estimation
forecasting are as follows \cite{MPCLK}.
\begin{itemize}
\item Experiments motivated by model selection questions should be
quantified by their ability to answer such questions.
\item Data is simulated at each point in the parameter space, rather
than at only one or more fiducial models. Indeed in parameter estimation
plots people commonly simulate data for the model that they hope to
rule out, rather than for the true model that would allow that
exclusion. 
\item Model selection analyses can attribute positive support for a
simpler model, rather than only showing consistency.
\item Gaussian approximations to the likelihood are not made, such as 
in parameter estimation forecasting done using Fisher matrices.
\end{itemize}

In assessing the significance of a model comparison, a useful guide is
given by the Jeffreys' scale \cite{Jeff}. Labelling as $M_1$ the model
with the higher evidence, it rates \mbox{$\ln B_{10} < 1$} as `not
worth more than a bare mention', $1<\ln B_{10} < 2.5$ as
`substantial', $2.5< \ln B_{10} < 5$ `strong' to `very strong' and
$5<\ln B_{10}$ as `decisive'. Note that $\ln B_{10}=5$ corresponds to
odds of 1 in about 150, and $\ln B_{10}=2.5$ to odds of 1 in 13.

A model selection analysis of the Planck satellite's capabilities to
constrain $n_{{\rm S}}$ was previously given by Trotta \cite{T} using
his ExPO technique. This seeks to estimate the probability, based on
current knowledge of parameters, of the Planck mission being able to
carry out a decisive model comparison. Our aim is rather different; we
seek to delineate the parameter values the Universe would have to have
in order for a decisive model comparison to be made. However we will
end by additionally making an ExPO-style forecast, though with a
somewhat different implementation to Trotta's. 

Another related paper is Bridges et al.~\cite{BLH}, who simulate data
for a model with constant $n_{{\rm S}}$ and compare the evidences for
a set of initial power spectrum models. They do not however explore
different values of the spectral index.

It may seem strange that model selection approaches can give results
in apparent conflict with parameter estimation. However, this is a
well-known phenomenon called Lindley's paradox \cite{L,T}; the idea
that there is a universal significance level such as 95\% beyond which
things become interesting is inconsistent with Bayesian reasoning,
which shows that such a threshold should depend both on the data
properties and the prior parameter ranges. The Lindley paradox usually
manifests itself for results with significance in the range two to
four sigma \cite{T}, which as it happens is exactly where WMAP3 has
placed $n_{{\rm S}}$.

\subsection{Simulating Planck data}

In order to give a good estimate of Planck's abilities, we need
accurate data simulations. Simulated Planck data was generated by
Bridges et al.~\cite{BLH} for their model selection analysis, but they
simply assumed cosmic variance limited temperature anisotropies out to
$\ell = 2000$.\footnote{Trotta \cite{T} also used Planck simulations,
but did not disclose how they were implemented.}  We adopt a rather
more sophisticated approach, as follows.

We simulate the temperature and polarization (TT, TE, and EE) spectra. 
We choose not to include $B$-polarization for simplicity; as
we do not include tensors there are no primordial $B$ modes, and the
shorter-scale $B$-modes generated by gravitational lensing will not
supply significant constraining power on the specific models we are
considering.

We use three temperature channels, of specifications similar to the
HFI channels of frequency 100 GHz, 143 GHz, and 217 GHz. Following the
current Planck documentation,\footnote{
www.rssd.esa.int/index.php?project=PLANCK\&page=perf\underline{~}top}
the intensity sensitivities of these channels are taken as 6.8 $\mu$K,
6.0 $\mu$K, and 13.1 $\mu$K respectively, corresponding to the values
quoted for two complete sky surveys.  These are average sensitivities
per pixel, where a pixel is a square whose side is the FWHM extent of
the beam. The FWHM's of these channels are given as 9.5 arcmin, 7.1
arcmin, and 5.0 arcmin respectively.  The composite noise spectrum for
the three temperature channels is obtained by inverse variance
weighting the noise of individual channels \cite{K,CH}. For
polarization we take only one channel, the 143 GHz channel, of FWHM
7.1 arcmin, and sensitivity 11.5 $\mu$K.

The assumed Gaussianity of the spherical harmonic coefficients of the
temperature and polarization leads to a likelihood function given by
(see e.g.~Ref.~\cite{like})
\begin{equation}
-2 \log {\cal L} = (2{\ell}+1){\rm f_{sky}^2} \left[ {\rm Tr} 
\left( \hat{C_{\ell}} C_{\ell}^{-1} \right) + \log |C_{\ell}| \right]
\end{equation}
where 
\begin{equation}
C_{\ell} =  \left( \begin{array}{ccc}\vspace{1mm}
C_{\ell}^{TT} & C_{\ell}^{TE} & 0 \\ \vspace{1mm}
C_{\ell}^{TE} & C_{\ell}^{EE} & 0 \\ 
0 & 0 & C_{\ell}^{BB} \\ \end{array} \right)
\end{equation}
and $\hat{C_{\ell}}$ is the corresponding matrix of estimators.  Both
$C_{\ell}$ and $\hat{C_{\ell}}$ include instrumental noise variance.
The fractional sky covered is taken to be 0.8 for all $\ell$, and we
use simulated data out to an $\ell_{{\rm max}}$ of 2000.

We simulate data for a range of values of $n_{{\rm S}}$. In defining
the fiducial models for which the data are simulated, the other
parameters are kept fixed, those parameters being the cold dark matter
density $\Omega_{{\rm cdm}}$, the baryon density $\Omega_{{\rm B}}$,
the optical depth $\tau$, the angular size of the sound horizon at
decoupling $\Theta$ and the power spectrum amplitude $A_{{\rm
S}}$. The specific values chosen were $\Omega_{{\rm b}} h^2 = 0.024$,
$\Omega_{{\rm c}} h^2 = 0.103$, $\Theta = 1.047$, $\tau = 0.14$ and
$A_{{\rm S}}=2.3 \times 10^{-9}$ respectively, where $h$ is the Hubble
parameter in the usual units and is equal to $0.78$ for these
parameter choices. These values were motivated by the WMAP3 results
\cite{wmap3}. All parameters are varied in computing evidences.

\subsection{Results}

Having simulated Planck data for a given $n_{{\rm S}}$, we compute the
evidences of the two models, which we denote by HZ and VARYn. The
former is of course the Harrison--Zel'dovich model with $n_{{\rm S}}$
fixed to one. The latter is a model with $n_{{\rm S}}$ allowed to vary
in fits to the data. In each case, all the other parameters are
allowed to vary, each with the same prior range as used in
Ref.~\cite{MPL}. This is repeated for different values of fiducial
$n_{{\rm S}}$.

As in Ref.~\cite{MPL}, the prior range for $n_{{\rm S}}$ is taken to
be $0.8 < n_{{\rm S}} < 1.2$, representing a reasonable range allowed
by slow-roll inflation models (see e.g.~Ref.~\cite{LL}). The end
result does have some prior dependence. If the prior is widened in
regions where the likelihood is negligible, then the evidence just
changes proportional to the prior volume, so for instance a doubling
of the prior range will only reduce the ln(evidence) by $\ln 2 =
0.69$. This indicates that the prior range is not very important for
this parameter.

We use the CosmoNest algorithm described in Refs.~\cite{MPL,PML} to
compute the evidences. This is based on the nested sampling algorithm
of Skilling \cite{Skilling}, and is a fast Monte Carlo (but not Markov
chain) method for accurately averaging the likelihood across the
entire prior space. The algorithm parameters used were $N = 300$ live
points and an enlargement factor of 1.8 for HZ and 1.9 for
VARYn. These enlargement factors are higher than those required for
the same models and similar target accuracy with say WMAP data.  This
is because as the data improve the likelihood contours in the high
likelihood regions can deviate from elliptical and become more banana
shaped. The tolerance parameter was set to 0.5 which gave answers to
good accuracy as indicated by the uncertainties obtained. Four
independent evidence evaluations were done for each calculation, to
obtain the mean and its standard error.

\begin{figure}[t]
\centering
\includegraphics[width=8cm]{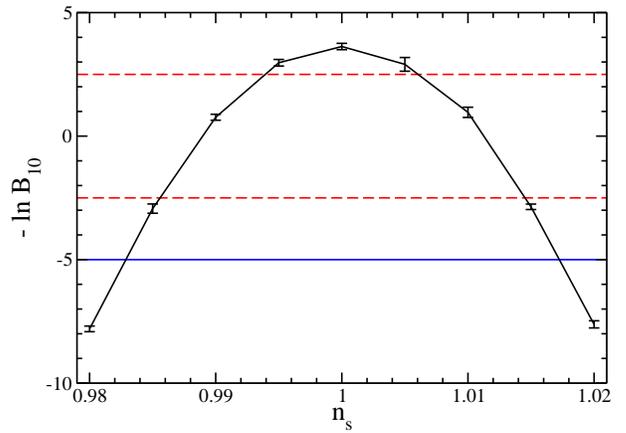}
\caption{\label{f:bayes} The (negative of the) logarithm of the Bayes
factor, $-\ln B_{10}$, as a function of the fiducial value of $n_{{\rm
S}}$, where $M_0$ is the HZ model and $M_1$ is VARYn. The horizontal
lines indicate where the comparison becomes `strong' (dashed) and
`decisive' (solid) on the Jeffreys' scale.}
\end{figure}

Figure~\ref{f:bayes} shows our main result. At $n_{{\rm S}} = 1$, the
HZ model is strongly preferred with $\ln B_{10}= - 3.6 \pm 0.1$. It
has a higher evidence since it can fit the data just as well as VARYn
and has one less parameter. Once $n_{{\rm S}}$ is far enough away from
1, the HZ fit becomes very poor and the Bayes factor plummets. The
speed with which this happens indicates the strength of the
experiment.

We see that if the true value lies in the range $0.989 < n_{{\rm S}}
<1.011$, Bayesian model selection will favour the HZ model, and within
the narrower range $0.994 < n_{{\rm S}} <1.006$ it will give strong
support to that model, though Planck on its own is not powerful enough
to be able to decisively favour HZ over VARYn even if HZ is the true
case. Only once $n_{{\rm S}} < 0.986$ or $n_{{\rm S}} > 1.014$ can
Planck offer strong evidence against HZ, rapidly becoming decisive as
the fiducial value moves away from unity beyond 0.983 or 1.017.

We can contrast these model selection results with those indicated by
parameter estimation. Using the same simulated data, we compute the
marginalized likelihood of $n_{{\rm S}}$ about $n_{{\rm S}}=1$. This
gives a 68\% range of $0.995 < n_{{\rm S}} < 1.004$ and a 95\% range
of $0.991 < n_{{\rm S}} < 1.008$, in good agreement with estimates
obtained by other authors including the Planck Blue Book. We see this
is an explicit example of Lindley's paradox; there are values of
$n_{{\rm S}}$ lying outside the 95 percent confidence region, for
which model selection would nevertheless favour the HZ model.

We end by estimating how likely it is that Planck will be able to make
a decisive selection between our two models, based on current
understanding of the spectral index. We use a variant of Trotta's ExPO
approach \cite{T}, but with one important distinction; that we use the
current model selection position as input, whereas Trotta used the
observed likelihood in the VARYn model alone. For simplicity we
consider only the marginalized likelihood for $n_{{\rm S}}$ as the
starting point rather than marginalizing the model selection outcome
over all parameters, but we expect that to make little difference in
this case.

According to Parkinson et al.~\cite{PML}, following WMAP3 the balance
of probability between HZ and VARYn is 12\% to 88\% (with some
dependence on the choice of data compilation). This makes the
important assumption that the models were thought equally likely
before the data came along; anyone who thinks otherwise can readily
recompute according to their own prejudice. In essence, one can think
of the probability distribution for $n_{{\rm S}}$ as being a weighted
superposition of the likelihood in the VARYn model plus a
delta-function at $n_{{\rm S}} = 1$. Trotta omits the delta-function
term in his ExPO forecasts.

For the 12\% probability that $n_{{S}}$ is actually one, Planck will
clearly not find evidence to the contrary, but as we have seen would
in that case provide strong evidence {\em for} the HZ case. For the
remaining probability, we use the marginalized distribution for
$n_{{\rm S}}$ as computed in Ref.~\cite{PML}. We find that 7\% of the
posterior lies in the region $n_{{\rm S}} > 0.983$ where even Planck
cannot make a decisive verdict.  We can therefore conclude that if
$n_{{\rm S}}$ is not one, then Planck is expected to provide a
decisive verdict against HZ, which WMAP3 has not achieved, but with a
small chance it will not.

Trotta \cite{T} came to the same verdict that Planck is very likely to
rule out $n_{{\rm S}} = 1$, but using WMAP1 data. However we would
{\em not} have come to that conclusion using our modification of his
approach, as with WMAP1 the model selection verdict put somewhat more
than half the probability in the HZ case \cite{MPL}, and also a
significant part of the $n_{{\rm S}} \neq 1$ probability into the
indecisive region. Accordingly, at that point we would have said that
the most likely outcome of Planck (under the assumption of equal model
prior probabilities) was strong support {\em for} $n_{{\rm S}} =
1$. However, as is always the danger with probabilities, WMAP3 has
overturned that conclusion.

\vspace*{50pt}

\section{Conclusions}

We have carried out a model selection forecast for the Planck
satellite, focussing on the scalar spectral index. Such analyses
complement the usual parameter error forecasts, and are particularly
directed to the question of when one can robustly identify the need
for new fit parameters. In particular, we have delineated the values
of $n_{{\rm S}}$ for which strong or decisive model comparisons can be
carried out. Ruling out $n_{{\rm S}}=1$ is found to be significantly
harder than parameter error forecasts suggest.

The recent WMAP3 data have left $n_{{\rm S}}$ poised in an interesting
position, where model selection analyses do support parameter
estimation conclusions but not yet at a decisive level.  Our results
show that if $n_{{\rm S}}$ really is different from one, then Planck is
very likely to be able to confirm that, but if the HZ case
is the true one then even Planck will not be decisive.

\begin{acknowledgments}
A.R.L., P.M. and D.P. were supported by PPARC.  We thank Anthony
Challinor, Pier Stefano Corasaniti, Martin Kunz, Antony Lewis, and
Roberto Trotta for helpful discussions relating to this project.  The
CosmoNest code has been written as an add-on to CosmoMC, so we
acknowledge the use of CosmoMC.
\end{acknowledgments}


\end{document}